\documentclass{elsart}
\usepackage{graphicx,amsmath,amsfonts, trees, ecltree, epic, amsmath, amssymb, amsfonts}
\def\bgamma{{\mbox{\boldmath$\gamma$}}}
\begin{document}
\begin{frontmatter}

\title{Time evolution of the relativistic unstable electromagnetic system
in the unified formulation of quantum and kinetic dynamics}
\author{S.Eh.Shirmovsky}
\ead{shirmov@ifit.phys.dvgu.ru} \address{ Laboratory of
Theoretical Nuclear Physics, Far Eastern National University,
Sukhanov Str. 8, Vladivostok, 690950, Russia}
\date{\today}

\begin{abstract}
Description of time evolution of the relativistic unstable
electromagnetic system consisting of Fermi~-~Dirac particle
interacting with electromagnetic field, in the framework of the
Liouville space extension of quantum mechanics is done. The work
was carried out on the basis of Prigogine's unified formulation of
quantum and kinetic dynamics. The eigenvalues problem for the
relativistic Hamiltonian of the electromagnetic system was solved.
The obtained results can be used as the ground for the further
studies of the observed physical processes such as bremsstrahlung,
relaxation of excited states of atoms and atomic nuclei,
particles decay.
\end{abstract}
\begin{keyword}bremsstrahlung, electromagnetic, irreversibility, nonequilibrium
\PACS{12.20.-m, 03.65.-w, 13.40.Hq}
\end{keyword}
\end{frontmatter}

\section{Introduction}

It is known that the description of physical world on the basis of
fundamental classical and quantum theories is defined by the laws
of the nature as deterministic, time reversible. The time in the
usual formulation of dynamics does not have the chosen direction
and the future, and the past are not distinguished. However, it is
also noted that the facts given before are in the contradiction to
our experience, because the world surrounding us has obvious
irreversible nature. In this world the symmetry in the time is
disrupted and the future and the past play different roles.
Difference between the classical description of the nature ¶and
those processes in the nature which we observe creates the
conflict situation. I. Prigogine noted that the solution of this
problem is impossible (at least accurately) on the basis of the
conventional formulation of quantum mechanics. Therefore one
should speak about the alternative formulation of dynamics that
makes it possible to include the irreversibility in a natural way.
In this connection the studies of the irreversible processes at
the microscopic level - the microscopic formulation of the
irreversibility represents for me the special interest. \\
The alternative formulation of dynamics found its embodiment in
the works of Brussels-Austin group that was headed by I. Prigogine
for many years. The authors of the approach deny the conventional
opinion that the irreversibility appears only at the macroscopic
level, while the microscopic level must be described by the laws,
reversed in the time. The mechanism of the asymmetry of processes
in the time, which made it possible to accomplish a passage from
the reversible dynamics to the irreversible time evolution one was
developed. In the approach the law of the increase of entropy is
accepted as the fundamental that determines the "arrow of time",
the difference between the past and the future. Thus, new
irreversible dynamics with the disrupted symmetry in the time was
formulated. In the approach of Brussels-Austin group the
¶irreversibility is presented as the property of material itself
and is not defined by the active role of the observer. From the
other side the approach allows to solve the problems, which could
not be solved within the framework of classical and quantum
mechanics. For example, now we can realize the program of
Heisenberg - to solve the eigenvalues problem for the Poincare's
non-integrable systems, which could not be solved within the
framework of traditional methods. The approach solves the basic
problem, designated by Boltzmann and Planck - to formulate the
second law of thermodynamics at the microscopic level.\\
The general formalism of Brussels-Austin group approach was
developed in works~\cite{p}~-~\cite{pp8}. In these papers the
basic ideas of the transition from deterministic dynamics to the
irreversible description are formulated. In
monographs~\cite{p},~\cite{ps} the universal survey of ideas and
principles of the alternative formulation of dynamics is given. A
unified formulation of dynamics and thermodynamics is done, for
example, in works~\cite{pghr},~\cite{mpc}. It is carried out the
study of the non-integrable systems, where the alternative
formulation of quantum mechanics for non-integrable systems is
proposed~\cite{tp},~\cite{ppt}. The "subdynamics" approach is
developed in works~\cite{pgh4},~\cite{pt},~\cite{ph},~\cite{pp3}.
In paper~\cite{at} the method was adapted for the explicit
computation of the eigenvalues problem for the Renyi maps, baker's
transformations, Fridrichs model. The eigenvalues problem for the
Liouville operator $L$ is solved in the framework of a complex,
irreducible spectral representation~\cite{pp9},~\cite{pp10}. The
role of the thermodynamic limit for Large Poincare systems is
investigated in work \cite{pp8}. In works~\cite{gmp},~\cite{pom},
scattering theory in superspace and the three-body scattering
theory for finite times is developed. In~\cite{opp}~-~\cite{kop2}
in the framework of Friedrichs model the problem of description of
quantum unstable states including their dressing is investigated.
The problem of the complex spectral representation of Liouville
operator in the extended Liouville space outside the Hilbert space
is solved. It is shown that the dressed unstable state described
by a density matrix can be expressed in the terms of the Gamow
vectors. The Gamow vectors are investigated also in
works~\cite{agpp}~-~\cite{agkpp342}. The formalism determines the
operator of microscopic entropy $M$ and also the time operator
$T$~\cite{opkp}. The time operator $T$ is constructed for a
quantum system with unstable particle. In work~\cite{po}
one-dimensional gas with $\delta$-function interaction is
examined. Formalism found its further development in
works~\cite{kopp}~-~\cite{ptb2}. So in work~\cite{kopp} on basis
of Friedrichs model a microscopic expression for entropy is
obtained. The simple model of interaction harmonic oscillator with
a field is developed in work~\cite{pop4}. A problem of quantum
decoherence for a particle coupled with a field was considered in
work~\cite{ptb}. It was shown that the decoherence in the field is
the result of irreversible process. In work~\cite{akp2}  analysis
of the Hegerfeldt's theorem is carried out. The analysis of the
short-time behaviour of the survival probability in the framework
of the Friedrichs model was done in work~\cite{akpy}. The two
models of relativistic interaction are examined in the
work~\cite{agmp}. The models involve two relativistic quantum
fields. They are coupled by the simplest cubic and quadratic
interaction.  A pair of identical two-level atoms interacting with
a scalar field are considered in the work~\cite{ptb2}. The
questions of irreversibility are developed also in
works~\cite{g3}~-~\cite{hd2}.\\
At present in the framework of Prigogine's ideas the great number
of works with the use of different models of interaction was
executed. They are the Friedrichs model
(see~\cite{ppp}~-~\cite{am}), the models used interaction of
simple cubic, or quadratic form as, for example, in
works~\cite{agpp},~\cite{agmp} or $\delta$-function
interaction~\cite{po}. Therefore, at present moment, it is very
interesting and necessary to continue further development of the
formalism with the use of realistic relativistic Hamiltonians.
\\
In the paper I examine the possibility of application
Brussels-Austin group's ideas, for the description of time
irreversible evolution of the quantum  unstable electromagnetic
system consisting of Fermi~-~Dirac particle interacting with
electromagnetic field. The model of relaxation of the system with
photon emission is investigated. It is important that the
Hamiltonian of interaction is determined on the basis of the
requirement of the gauge invariance of the model. The definition
of the interaction model is done in section 2. In section 3 I
present the Liouville formalism, "subdynamics" approach. In
section 4 the initial expression for the density matrix is
formulated. The task of the complex spectral representation of
Hamiltonian is solved in section 5. The expression for the density
matrix describing the evolution of the relativistic, unstable
electromagnetic system depending on the time is obtained in
section 6. Numerical calculation are given in section 7.

\section{Definition of the interaction model}

For the operator of fermion field $\psi(x)$ we have the following
decomposition~\cite{Bil}
\begin{equation}\label{psidecoper50}
\begin{split}
&\psi(x)=\psi^{(+)}(x)+\psi^{(-)}(x),\\
\psi^{(+)}(x)&=\frac{1}{(2\pi)^{3/2}}\int
\Bigl{(}\frac{m}{p_{0}}\Bigr{)}^{1/2}u^{r}(p)e^{ipx}c_{r}(p)d\textbf{p}, \\
\psi^{(-)}(x)&=\frac{1}{(2\pi)^{3/2}}\int
\Bigl{(}\frac{m}{p_{0}}\Bigr{)}^{1/2}u^{r}(-p)e^{-ipx}d^{\dag}_{r}(p)d\textbf{p},
\end{split}
\end{equation}
where $c_{r}(p)$ ($c^{\dag}_{r}(p)$) is the operator of
destruction (creation) of the particle, $d^{\dag}_{r}(p)$
($d_{r}(p)$) is the operator of creation (destruction) of
antiparticle. Symbol $"\dag"$ indicates the Hermitian conjugate.
Operator $\psi(x)$ satisfies the Dirac equation and evolves
according to the Dirac representation, satisfying the expression
\begin{align}\label{repint}
i\frac{\partial \psi(x)}{\partial t}=[\psi (x), H_{0}].
\end{align}
$H_{0}$ in eq.~\eqref{repint} is a free Hamiltonian. Note that we
will write 4~- vectors in the form $A=(\textbf{A}, iA_{0})$. In
this case the following equalities are valid
$A^{2}=\textbf{A}^{2}+A_{4}^{2}=\textbf{A}^{2}-A^{2}_{0}$ and
$px=p_{\mu}x_{\mu}=\textbf{px}-p_{0}x_{0}$;
$p_{0}=\sqrt{\textbf{p}^{2}+m^{2}}$, $m$ - is mass of the quantum
of field. We use units with $\hbar$, and the speed of light $c$
taken to be unity ($\hbar = c = 1$). ¶ Spinors $u^{r}(p)$,
$u^{r}(-p)$ correspond to the particles with helicity $r=\pm 1$. \\
We determine the operator of electromagnetic field as follows
\begin{equation}\label{ae}
\begin{split}
&A_{\mu}(x)=A^{(+)}_{\mu}(x) + A^{(-)}_{\mu}(x),\\
A^{(+)}_{\mu}(x)&= \frac{1}{(2\pi)^{3/2}}\int
\frac{1}{\sqrt{2k_{0}}}\sum\limits_{\lambda}e^{\lambda}_{\mu}(k)a_{\lambda}(k)e^{ikx}d\textbf{k},\\
A^{(-)}_{\mu}(x)&= \frac{1}{(2\pi)^{3/2}}\int
\frac{1}{\sqrt{2k_{0}}}\sum\limits_{\lambda}e^{\lambda}_{\mu}(k)a^{\dag}_{\lambda}(k)e^{-ikx}d\textbf{k},
\end{split}
\end{equation}
where $k_{0}=|\textbf{k}|$, $a_{\lambda}(k)$
($a^{\dag}_{\lambda}(k)$) is the operator of destruction
(creation) of $\gamma$-quantum. Operators of the electromagnetic
field $A_{\mu}(x)$ evolve according to the Dirac representation as
well. Polarization vector $e^{\lambda}_{\mu}(k)$ is determined by
the relations
\begin{align}\label{aexpr}
e^{\lambda}(k)=(\textbf{e}^{\lambda}(\textbf{k}), 0),
~e^{4}(k)=(\textbf{0},i),~\lambda =1, 2, 3~(\text{polarization
index}).
\end{align}
$\textbf{e}^{1}(\textbf{k})$, $\textbf{e}^{2}(\textbf{k})$ are
unit vectors orthogonal to each other and to the momentum of
$\gamma$ - quantum $\textbf{k}$, $\textbf{e}^{3}(\textbf{k})$ is
unit vector directed along vector $\textbf{k}$. \\
On the basis of the requirement of gauge invariance of the model
the Hamiltonian of interaction must be determined in the
conventional form~\cite{Bil} (see also~\cite{shv})
\begin{align}\label{ae90}
H_{I}(t) = -ie\int
N(\overline{\psi}(x)\gamma_{\mu}\psi(x))A_{\mu}(x)d\textbf{x}.
\end{align}
$\gamma_{\mu}$- Hermitian 4$\times$4 matrices
($\gamma_{\mu}$$\gamma_{\nu}$ + $\gamma_{\nu}$$\gamma_{\mu}$ =
2$\delta_{\mu\nu}$, $\gamma^{\dag}_{\mu}=\gamma_{\mu}$),
$\overline{\psi}$ = $\psi^{\dag}$$\gamma_{4}$, $N$ is the symbol
of the normal ordering of operators, $e$ is the charge of the
electron so that the fine structure constant is:
$\alpha=e^{2}/4\pi\simeq 1/137$.

\section{Liouville formalism, "subdynamics" }

Now let me briefly examine the Liouville formalism (see for
example~\cite{opp},~\cite{kop2}). The time evolution of the
density matrix $\rho$ is determined by the Liouville-von Neumann
equation
\begin{equation}\label{ter}
i\frac{\partial\rho}{\partial t}= L\rho.
\end{equation}
Liouville operator has the form
\begin{align}\label{liuvil}
L=H\times 1 - 1\times H,
\end{align}
here symbol "$\times$" denotes the operation
(A$\times$B)$\rho$=A$\rho$B. In accordance with
formula~\eqref{liuvil}, $L$ can be written down in the sum of free
part $L_{0}$ that depends on the free Hamiltonian $H_{0}$ and
interaction part $L_{I}$ that depends on $H_{I}$:  $L=L_{0} +
L_{I}$. Let state $|\alpha\rangle$ be the eigenstate of the free
Hamiltonian $H_{0}|\alpha\rangle$=$E_{\alpha}|\alpha\rangle$ with
the energy $E_{\alpha}$. Then dyad of the states
$|\alpha\rangle\langle\beta|$ is the eigenstate of operator
$L_{0}$:
$L_{0}|\alpha\rangle\langle\beta|=(E_{\alpha}-E_{\beta})|\alpha\rangle\langle\beta|$
or $L_{0}|\nu\rangle\rangle=w^{\nu}|\nu\rangle\rangle$, where the
designations $|\nu\rangle\rangle\equiv|\alpha\rangle\langle\beta|$
and $w^{\nu}=E_{\alpha}-E_{\beta}$  were used. $\nu$ is the
correlation index: $\nu = 0$ if $\alpha=\beta$ - diagonal case
(vacuum of correlation) and $\nu\neq 0$ in the remaining
off-diagonal case (the details of the theory of correlations can
be found in works~\cite{ps},~\cite{ppt},~\cite{opp}). In the
Liouville space for the dyadic operators we have the relations:
inner product defined by (where symbol $Tr$ denotes the
calculation of the trace)
\begin{align}\label{j}
\langle\langle A|B\rangle\rangle \equiv Tr(A^{\dag}B),
\end{align}
the matrix elements are given by
\begin{align}\label{j2}
\langle\langle \alpha \beta|A\rangle\rangle\equiv\langle
\alpha|A|\beta\rangle,
\end{align}
the biorthogonality and bicompleteness relations have the form
\begin{align}\label{j3}
\langle\langle \alpha^{'}\beta^{'}|\alpha \beta\rangle\rangle =
\delta_{\alpha' \alpha}\delta_{\beta'
\beta},~\sum\limits_{\alpha,\beta}|\alpha\beta\rangle\rangle\langle\langle\alpha\beta|=1.
\end{align}
It was shown that the description of the irreversible processes at
the microscopic level is possible if eigenvalues of Liouvillian
$Z^{\nu}_{j}$ are generally complex. Thus, for the Liouville
operator $L$ the eigenvalues problem is formulated for the
right-eigenstates $|\Psi^{\nu}_{j}\rangle \rangle$ and for the
left-eigenstates $\langle \langle
\widetilde{\Psi}^{\nu}_{j}|$~\cite{opp}
\begin{align}\label{ae4}
L|\Psi^{\nu}_{j}\rangle \rangle =Z^{\nu}_{j}|\Psi^{\nu}_{j}\rangle
\rangle ,~\langle \langle \widetilde{\Psi}^{\nu}_{j}|L=\langle
\langle \widetilde{\Psi}^{\nu}_{j}|Z^{\nu}_{j}.
\end{align}
Since $L$ is Hermitian the eigenstates are outside the Hilbert
space~\cite{kop2}. In this case the corresponding eigenstates have
no Hilbert norm~\cite{pp3},~\cite{opp}. For
$|\Psi^{\nu}_{j}\rangle \rangle$ and $\langle \langle
\widetilde{\Psi}^{\nu}_{j}|$ we have the following biorthogonality
and bicompleteness relations
\begin{align}\label{ae5}
\langle\langle\widetilde{\Psi}^\nu_{j}|\Psi^{\mu}_{j'}
\rangle\rangle= \delta_{\nu\mu}\delta_{j j'},
~\sum\limits_{\nu,j}|\Psi^{\nu}_{j}\rangle\rangle
\langle\langle\widetilde{\Psi}^\nu_{j}|=1.
\end{align}
Index $j$ is a degeneracy index since one type of correlation
$\nu$ can correspond to different states.\\
It is shown in work~\cite{opp} that the eigenstates of $L$ can be
written in terms of kinetic operators $C^{\nu}$ and
$D^{\nu}$~\cite{crit},~\cite{crit2}. Operator $C^{\nu}$ creates
correlations other than the $\nu$ correlations, $D^{\nu}$ is
destruction operator. The use of the kinetic operators allows to
write down the expressions for the eigenstates of Liouville
operator in the following form
\begin{align}\label{ae6}
|\Psi^{\nu}_{j}\rangle\rangle=(N^{\nu}_{j})^{1/2}\Phi^{\nu}_{C}|u^{\nu}_{j}\rangle\rangle
,~\langle\langle\widetilde{\Psi}^{\nu}_{j}|=\langle\langle
\widetilde{v}^{\nu}_{j}|\Phi^{\nu}_{D}(N^{\nu}_{j})^{1/2},
\end{align}
where
\begin{align}\label{ae7}
\Phi^{\nu}_{C}\equiv P^{\nu} + C^{\nu},~\Phi^{\nu}_{D}\equiv
P^{\nu} + D^{\nu}
\end{align}
and $N^{\nu}_{j}$ - is a normalization constant. In the general
case the operators $P^{\nu}$ satisfy the following
condition~\cite{pp3}
\begin{align}\label{ae8}
P^{\nu}=\sum\limits_{j}|u^{\nu}_{j}\rangle\rangle
\langle\langle\widetilde{u}^{\nu}_{j}|,
~\langle\langle\widetilde{u}^{\nu}_{j}|u^{\mu}_{j'}\rangle\rangle=\delta_{\nu
\mu}\delta_{j j'}.
\end{align}
Similarly for $P^{\nu}$ and
$\langle\langle\widetilde{v}^{\nu}_{j}|$ we have
\begin{align}\label{ae8}
P^{\nu}=\sum\limits_{j}|v^{\nu}_{j}\rangle\rangle
\langle\langle\widetilde{v}^{\nu}_{j}|,
~\langle\langle\widetilde{v}^{\nu}_{j}|v^{\mu}_{j'}\rangle\rangle=\delta_{\nu
\mu}\delta_{j j'}.
\end{align}
The determination of the states $|u^{\nu}_{j}\rangle\rangle$,
$\langle\langle\widetilde{v}^{\nu}_{j}|$ can be found in
work~\cite{opp}. Substituting~\eqref{ae6} in~\eqref{ae4} and
multiplying $P^{\nu}$ from left on both sides, we
obtain~\cite{opp}
\begin{align}\label{ae8}
\theta^{\nu}_{C}|u^{\nu}_{j}\rangle\rangle =
Z^{\nu}_{j}|u^{\nu}_{j}\rangle\rangle,
\end{align}
where
\begin{align}\label{ae9}
\theta^{\nu}_{C}\equiv P^{\nu}L\Phi^{\nu}_{C}=L_{0}P^{\nu} +
P^{\nu}L_{I}\Phi^{\nu}_{C}P^{\nu}.
\end{align}
$\theta^{\nu}_{C}$ is the collision operator connected with the
kinetic operator $C^{\nu}$. This is non-Hermitian dissipative
operator, which plays a main role in nonequilibrium dynamics. As
was shown in ref.~\cite{pp3} the case $\nu = 0$ leads
$\theta^{0}_{C}$ to the collision operator in the Pauli master
equation for weakly coupled systems.\\
Analogously it is possible to obtain equation for operator
$\theta^{\nu}_{D}$, which is connected with the destruction
kinetic operator $D^{\nu}$.¶
\begin{align}\label{tetade}
\langle\langle\widetilde{v}^{\nu}_{j}|\theta^{\nu}_{D}=\langle\langle\widetilde{v}^{\nu}_{j}|Z^{\nu}_{j},
\end{align}
where
\begin{align}\label{tetad}
\theta^{\nu}_{D}\equiv L_{0}P^{\nu} +
P^{\nu}\Phi^{\nu}_{D}L_{I}P^{\nu}.
\end{align}
Comparing eqs.~\eqref{ae4},~\eqref{ae8},~\eqref{tetade}  we can
see that $|u^{\nu}_{j}\rangle\rangle$ and
$\langle\langle\widetilde{v}^{\nu}_{j}|$ are eigenstates of
collision operator $\theta^{\nu}_{C(D)}$ with the same
eigenvalues $Z^{\nu}_{j}$ as $L$.\\
Thus, determination of the eigenvalues problem for the Liouville
operator $L$ outside the Hilbert space leads to the
connection of quantum mechanics with kinetic dynamics.\\
Operators $\Phi^{\nu}_{C}$, $\Phi^{\nu}_{D}$ satisfy so-called
"nonlinear Lippmann-Schwinger equation". For the $\Phi^{\nu}_{C}$
we have~\cite{opp}
\begin{align}\label{fieq}
\Phi^{\nu}_{C}=P^{\nu}+\sum\limits_{\mu\neq\nu}
P^{\mu}\frac{-1}{w^{\mu}-w^{\nu}-i\varepsilon_{\mu\nu}}[L_{I}\Phi^{\nu}_{C}-\Phi^{\nu}_{C}L_{I}\Phi^{\nu}_{C}]P^{\nu},
\end{align}
where the time ordering is introduced. For the determination of
the sign of the infinitesimals $\varepsilon_{\mu\nu}$ it is
necessary to determine the degree of correlation $d_{\mu (\nu)}$.
This was defined as the minimum number of interactions $L_{I}$ by
which a given state can reach the vacuum of correlation. It is
assumed that the directions to the higher degrees of correlation
are oriented in the future, and the directions to the lowest
degrees of correlation are oriented in the past. This leads to the
relations~\cite{opp},~\cite{georg}
\begin{align}\label{tetad}
\varepsilon_{\mu\nu}=+\varepsilon~ \text{if}~d_{\mu}\geq
d_{\nu}~(t>0);~\varepsilon_{\mu\nu}=-\varepsilon~
\text{if}~d_{\mu}< d_{\nu}~(t<0).
\end{align}
For the $\Phi^{\nu}_{D}$ we have the equation
\begin{align}\label{fieq2}
\Phi^{\nu}_{D}=P^{\nu}+P^{\nu}[\Phi^{\nu}_{D}L_{I}-\Phi^{\nu}_{D}L_{I}\Phi^{\nu}_{D}]
\sum\limits_{\mu\neq\nu}
P^{\mu}\frac{1}{w^{\nu}-w^{\mu}-i\varepsilon_{\nu\mu}}.
\end{align}
Eqs.~\eqref{fieq},~\eqref{fieq2} determine the kinetic operators
of creation $C^{\nu}$ and destruction $D^{\nu}$ as follows
\begin{equation}\label{cdeqq}
 \begin{split}
C^{\nu}&=\sum\limits_{\mu\neq\nu}
P^{\mu}\frac{-1}{w^{\mu}-w^{\nu}-i\varepsilon_{\mu\nu}}[L_{I}\Phi^{\nu}_{C}-\Phi^{\nu}_{C}L_{I}\Phi^{\nu}_{C}]P^{\nu},\\
D^{\nu}&=P^{\nu}[\Phi^{\nu}_{D}L_{I}-\Phi^{\nu}_{D}L_{I}\Phi^{\nu}_{D}]
\sum\limits_{\mu\neq\nu}
P^{\mu}\frac{1}{w^{\nu}-w^{\mu}-i\varepsilon_{\nu\mu}}.
\end{split}
\end{equation}

The spectral representation of the Liouville operator can be
written down in the form
\begin{align}\label{cdeq}
L=\sum\limits_{\nu,j}Z^{\nu}_{j}|\Psi^{\nu}_{j}\rangle\rangle\langle\langle\widetilde{\Psi}^{\nu}_{j}|
.
\end{align}
In the Brussels-Austin group approach "subdynamics" is called the
construction of a complete set of spectral projectors $\Pi^{\nu}$
~\cite{ph}, \cite{g3}, \cite{hgm7}
\begin{align}\label{poper}
\Pi^{\nu}=\sum\limits_{j}|\Psi^{\nu}_{j}\rangle\rangle\langle\langle\widetilde{\Psi}^{\nu}_{j}|.
\end{align}
The projectors $\Pi^{\nu}$ satisfy the following relations
\begin{equation}\label{com}
\begin{split}
&\Pi^{\nu}L=L\Pi^{\nu},~(\text{commutativity});~
\sum\limits_{\nu}\Pi^{\nu}=1,~(\text{completeness});\\
&\Pi^{\nu}\Pi^{\nu'}=\Pi^{\nu}\delta_{\nu\nu'},~(\text{orthogonality});
~\Pi^{\nu}=(\Pi^{\nu})^{\ast},~(\text{star-Hermiticity}),
\end{split}
\end{equation}
where the action "$\ast$" corresponds to the "star" conjugation,
which is Hermitian conjugation plus the change
$\varepsilon_{\mu\nu}\rightarrow
\varepsilon_{\nu\mu}$~\cite{ppt}~\cite{pp3}. Oprerator $\Pi^{\nu}$
can be represented in the following form~\cite{opp}
\begin{align}\label{pca3}
\Pi^{\nu}=(P^{\nu}+C^{\nu})A^{\nu}(P^{\nu}+D^{\nu}),
\end{align}
where $A^{\nu}$ is the star-Hermitian operator
\begin{align}\label{pca2}
A^{\nu}=P^{\nu}(P^{\nu}+D^{\nu}C^{\nu})^{-1}P^{\nu}.
\end{align}
Taking~\eqref{com} it is possible to write down the density matrix
$\rho$ as follows
\begin{align}\label{poper}
\rho=\sum\limits_{\nu}\Pi^{\nu}\rho=\sum\limits_{\nu}\rho^{\nu},
\end{align}
where $\rho^{\nu}=\Pi^{\nu}\rho$. Projectors $\Pi^{\nu}$ can be
associated with the introduction of the concept of "subdynamics"
because the components $\rho^{\nu}$ satisfy separate equations. In
the framework of the "subdynamics" approach the time evolution of
the density matrix has the form (see, for example,
work~\cite{pp3})
\begin{equation}\label{ffi}
\begin{split}
&\rho^{\nu}(t)\equiv\Pi^{\nu}\rho(t)=exp(-iLt)\Pi^{\nu}\rho(0)=\\
&=(P^{\nu}+C^{\nu})e^{-i\theta^{\nu}_{C}t}A^{\nu}(P^{\nu}+D^{\nu})\rho(0).
\end{split}
\end{equation}

\section{Time evolution of the density matrix }

Multiplying $\Pi^{\nu}$ and $P^{\nu}$ from the left on both sides
of eq.~\eqref{ter} we obtain
\begin{align}\label{poper5}
i\frac{\partial}{\partial t}P^{\nu}\rho^{\nu}=P^{\nu}L\rho^{\nu}.
\end{align}
Let operators $Q^{\nu}$ determine subspace ortogonal $P^{\nu}$.
The states belonging ¶to subspace $Q^{\nu}$ have a degree of
correlation differing from those which have the states belonging
to subspace $P^{\nu}$
\begin{align}\label{poper2}
P^{\nu}Q^{\nu}=Q^{\nu}P^{\nu}=0,
\end{align}
\begin{equation}\label{pqu}
~P^{\nu}+Q^{\nu}=1.
\end{equation}
Using oprerators $P^{\nu}$ and $Q^{\nu}$ we can rewrite
eqs.~\eqref{cdeqq}: $C^{\nu}=Q^{\nu}C^{\nu}P^{\nu}$,
$D^{\nu}=P^{\nu}D^{\nu}Q^{\nu}$. It is easy to see that operators
$C^{\nu}$ describe transitions from $P^{\nu}$ correlation subspace
to the $Q^{\nu}$ correlation subspace and operators $D^{\nu}$
describe transitions from the correlation subspaces other than
$\nu$ to the $\nu$ subspace. The
expressions~\eqref{pca3},~\eqref{poper5} and~\eqref{pqu} result
into~\cite{opp}
\begin{align}\label{kineq}
i\frac{\partial}{\partial
t}P^{\nu}\rho^{\nu}=\theta^{\nu}_{C}P^{\nu}\rho^{\nu}.
\end{align}
The case $\nu$=0 leads~\eqref{kineq} to the kinetic Pauli master
equation for $P^{\nu}\rho^{\nu}$ - component. Eq.~\eqref{kineq}
describes the time irreversible evolution of the unstable state.
Our great interest is to investigate eq.~\eqref{kineq} for the
system of interacting relativistic Fermi~-~Dirac particle and
electromagnetic field. For this purpose we will obtain the
Liouville - von Neumann equation for $P^{\nu}\rho^{\nu}$-
component in the Dirac representation. In the Dirac representation
for eq.~\eqref{kineq} we have
\begin{align}\label{Dirac}
i\frac{\partial }{\partial
t}P^{\nu}\rho^{\nu}(t)=P^{\nu}L_{I}(t)C^{\nu}P^{\nu}[P^{\nu}\rho^{\nu}(t)].
\end{align}
Determining operator
\begin{align}\label{tet}
\vartheta^{\nu}(t)\equiv P^{\nu}L_{I}(t)C^{\nu}P^{\nu}
\end{align}
for eq.~\eqref{Dirac} we get
\begin{align}\label{Diracequation}
i\frac{\partial}{\partial
t}P^{\nu}\rho^{\nu}(t)=\vartheta^{\nu}(t)P^{\nu}\rho^{\nu}(t),
\end{align}
where the previous designations of operators are preserved.\\
The general solution of eq.~\eqref{Diracequation} can be found
after examining the equivalent integral equation. The solution of
eq.~\eqref{Diracequation} we will search for the component
$P^{\nu}\rho^{\nu}(t)$
\begin{align}\label{Dt}
P^{\nu}\rho^{\nu}(t)=P^{\nu}\rho^{\nu}(t_{0})~
+~(-i)\int\limits^{t}_{t_{0}}dt_{1}\vartheta^{\nu}(t_{1})P^{\nu}\rho^{\nu}(t_{1}),
\end{align}
where $P^{\nu}\rho^{\nu}(t_{0})$ is the component into initial
$t=t_{0}$ moment of the time. Substituting in the right side of
the expression~\eqref{Dt} instead of $P^{\nu}\rho^{\nu}(t_{1})$
the sum
\begin{align}\label{Dts}
P^{\nu}\rho^{\nu}(t_{0})~
+~(-i)\int\limits^{t_{1}}_{t_{0}}dt_{2}\vartheta^{\nu}(t_{2})P^{\nu}\rho^{\nu}(t_{2})
\end{align}
and consecutively continuing this procedure we find  ¶
\begin{equation}\label{Dts}
\begin{split}
&P^{\nu}\rho^{\nu}(t)=\Bigl(1~+~
(-i)\int\limits^{t}_{t_{0}}dt_{1}\vartheta^{\nu}(t_{1})~+~(-i)^{2}\int\limits^{t}_{t_{0}}\int\limits^{t_{1}}_{t_{0}}
dt_{1}dt_{2}\vartheta^{\nu}(t_{1})\vartheta^{\nu}(t_{2})~+~\\
&...~+~(-i)^{n}\int\limits^{t}_{t_{0}}\int\limits^{t_{1}}_{t_{0}}~...~\int\limits^{t_{n-1}}_{t_{0}}
dt_{1}dt_{2}~...~dt_{n}\vartheta^{\nu}(t_{1})\vartheta^{\nu}(t_{2})~...~\vartheta^{\nu}(t_{n})~+~...
\Bigr )\times\\
&\times P^{\nu}\rho^{\nu}(t_{0}).
\end{split}
\end{equation}
The obtained expression can be written down in the form
\begin{align}\label{Dts4}
P^{\nu}\rho^{\nu}(t)=\Omega^{\nu}(t,~t_{0})P^{\nu}\rho^{\nu}(t_{0}),
\end{align}
where
\begin{align}\label{Dts5}
\Omega^{\nu}(t,~t_{0})=\sum\limits^{\infty}_{n=0}(-i)^{n}\int\limits^{t}_{t_{0}}\int\limits^{t_{1}}_{t_{0}}~...~\int\limits^{t_{n-1}}_{t_{0}}
dt_{1}dt_{2}~...~dt_{n}\vartheta^{\nu}(t_{1})\vartheta^{\nu}(t_{2})~...~\vartheta^{\nu}(t_{n}).
\end{align}
Operator $\Omega^{\nu}(t,~t_{0})$ in expression~\eqref{Dts4} plays
the role of the evolution operator. The non-Hermitian operator
$\Omega^{\nu}(t,~t_{0})$ determines the time irreversible
evolution of the density matrix - the time irreversible evolution
of the relativistic unstable state.

\section{Complex spectral representation of Hamiltonian}

Let me examine the eigenvalues problem for the  Hamiltonian
$H=H_{0}+H_{I}$. As for the Liouville  operator the problem will
be formulated outside the Hilbert space. In this case as earlier
we must distinguish equations for the right-eigenstates
$|\varphi_{\gamma}\rangle$ and for the left-eigenstates
$\langle\widetilde{\varphi}_{\gamma}|$ of Hamiltonian, where
$\gamma$ is the index of the state
\begin{align}\label{ht}
(H_{0} + H_{I})
|\varphi_{\gamma}\rangle=Z_{\gamma}|\varphi_{\gamma}\rangle,~\langle\widetilde{\varphi}_{\gamma}|(H_{0}
+ H_{I}) = \langle\widetilde{\varphi}_{\gamma}|Z_{\gamma},
\end{align}
eigenvalue $Z_{\gamma}$ is the complex number. Since $H$ is
Hermitian the corresponding eigenstates
$|\varphi_{\gamma}\rangle$, $\langle\widetilde{\varphi}_{\gamma}|$
have no Hilbert norm.
\begin{equation}
<\varphi_{\gamma}\mid\varphi_{\gamma}>=<{{\tilde{\varphi}}_{\gamma}}\mid{\tilde{\varphi}_{\gamma}}>=0.
\end{equation}
The "usual" norms of the states $|\varphi_{\gamma}\rangle$,
$\langle\widetilde{\varphi}_{\gamma}|$ disappear as required to
preserve the Hermiticity of $H$~\cite{ppt} (the details of the
complex eigenvalues problem can be found in
works~\cite{ps},~\cite{pop3}). Let me write down the Hamiltonian
of interaction in the form $H_{I}=gV$, determining explicitly
coupling constant $g$. The value of coupling constant depends on
the model of interaction and will be determined later. ¶Solutions
of the eqs.~\eqref{ht} can be found after presenting values
$|\varphi_{\gamma}\rangle$,
$\langle\widetilde{\varphi}_{\gamma}|$, $Z_{\gamma}$ in the
perturbation series
\begin{align}\label{ht3}
|\varphi_{\gamma}\rangle=\sum\limits_{n=0}^{\infty}g^{n}|\varphi_{\gamma}^{(n)}\rangle,~
\langle\widetilde{\varphi}_{\gamma}|=\sum\limits_{n=0}^{\infty}g^{n}\langle\widetilde{\varphi}_{\gamma}^{(n)}|,~
Z_{\gamma}=\sum\limits_{n=0}^{\infty}g^{n}Z_{\gamma}^{(n)},
\end{align}
where
\begin{align}\label{ht4}
|\varphi_{\gamma}^{(0)}\rangle=|\gamma\rangle,~
\langle\widetilde{\varphi}_{\gamma}^{(0)}|=\langle\gamma|,~
Z_{\gamma}^{(0)}=E_{\gamma}.
\end{align}
As it was shown in ref.~\cite{ppt} from
relations~\eqref{ht3},~\eqref{ht4} we can obtain
\begin{align}\label{ht5}
Z_{\gamma}^{(n)}=\langle\gamma|V|\varphi_{\gamma}^{(n-1)}\rangle~-~\sum\limits_{l=1}^{n-1}
Z_{\gamma}^{(l)}\langle\gamma|\varphi_{\gamma}^{(n-1)}\rangle,
\end{align}
\begin{align}\label{ht15}
\langle\beta|\varphi_{\gamma}^{(n)}\rangle=\frac{-1}{E_{\beta}-E_{\gamma}-i\varepsilon_{\beta\gamma}}
(\langle\beta|V|\varphi_{\gamma}^{(n-1)}\rangle~-~\sum\limits_{l=1}^{n}Z_{\gamma}^{(l)}\langle\beta|
\varphi_{\gamma}^{(n-1)}\rangle),
\end{align}
where in accordance with Brussels - Austin group approach the time
ordering is introduced. The sign of infinitesimal
$\varepsilon_{\beta\gamma}$ depends on the direction of the
processes: the transition $\gamma\rightarrow\beta$~we will
associate with
$\varepsilon_{\beta\gamma}=\varepsilon>0$. \\
Define $|\gamma\rangle$ as a bare state, which corresponds to
relativistic Fermi-Dirac particle and $|\beta\rangle$ as a state
consisting of the bare Fermi-Dirac particle and photon:
$|\gamma\rangle \equiv |\mathbf{p},r\rangle$, $|\beta\rangle
\equiv |\mathbf{p}',r';\mathbf{k},\lambda\rangle$, where
$|\mathbf{p},r\rangle$ refers to a one - particle state,
$|\mathbf{p}',r';\mathbf{k},\lambda\rangle$ is a two - particles
state, $\mathbf{p}$ $(\mathbf{p}')$, $r$ ($r'$) - momentum and
helicity of the particle and $\mathbf{k}$, $\lambda$ - momentum
and polarization index of photon. In the model, states
$|\mathbf{p},r\rangle$ $(|\mathbf{p}',r'\rangle)$,
$|\mathbf{k},\lambda\rangle$ are eigenstates of the free
Hamiltonian $H_{0}$: $H_{0}|\mathbf{p},r\rangle =
E_{\mathbf{p}}|\mathbf{p},r\rangle$,
$H_{0}|\mathbf{k},\lambda\rangle =
\omega_{\mathbf{k}}|\mathbf{k},\lambda\rangle$ with
$E_{\mathbf{p}}=\sqrt{\mathbf{p}^{2}+m^{2}}$ and
$\omega_{\mathbf{k}}=|\mathbf{k}|$ ($m$~-~Fermi-Dirac particle's
mass). Hamiltonian $H_{I}$ is determined by the
expression~\eqref{ae90}, $g\equiv e$ is the charge of the
electron. Substituting the expressions for
$\psi(x)$~\eqref{psidecoper50}, $A_{\mu}(x)$~\eqref{ae}
in~\eqref{ht5}, multiplying by $e^{n}$ and summing with respect to
$n$, we obtain
\begin{equation}\label{ht50}
\begin{split}
&Z_{\mathbf{p},r}=E_{\mathbf{p}}~-~ie
\frac{m}{(2\pi)^{3/2}}\int\sum\limits_{\lambda',r'}
\frac{d\mathbf{p}'d\mathbf{k}'}{(E_{\mathbf{p}}E_{\mathbf{p}'}2\omega_{\mathbf{k}'})^{1/2}}
~\delta(\mathbf{p}'-\mathbf{p}+\mathbf{k}')~\times~ \\
&\times
e^{i(E_{\mathbf{p}}-E_{\mathbf{p}'}-\omega_{\mathbf{k}'})t}
e^{\lambda'}_{\mu}(k')\overline{u}^{r}(p)\gamma_{\mu}u^{r'}(p')\frac
{\langle\mathbf{p}', r';~
\mathbf{k}',\lambda'|\varphi_{\mathbf{p}, r}\rangle}{\langle
\mathbf{p}, r|\varphi_{\mathbf{p},r}\rangle}.
\end{split}
\end{equation}
The expression for $\langle\mathbf{p}',
r';~\mathbf{k}',\lambda'|\varphi_{\mathbf{p}, r}\rangle$  can be
obtained from relation~\eqref{ht15}. In our case we get
\begin{equation}\label{ht20}
\begin{split}
&\langle\mathbf{p}', r';~\mathbf{k}',\lambda'|\varphi_{\mathbf{p},
r}\rangle = ie \frac{m}{(2\pi)^{3/2}}\int\sum\limits_{r''}
\frac{d\mathbf{p}''}{(E_{\mathbf{p}'}E_{\mathbf{p}''}2\omega_{\mathbf{k}'})^{1/2}}~\times\\
&\times\delta(\mathbf{p}''-\mathbf{p}'-\mathbf{k}')
e^{i(E_{\mathbf{p}'}-E_{\mathbf{p}''}+\omega_{\mathbf{k}'})t}
e^{\lambda'}_{\mu}(k')\overline{u}^{r'}(p')\gamma_{\mu}u^{r''}(p'')~\times\\
&\times\frac{\langle \mathbf{p}'', r''|\varphi_{\mathbf{p}, r
}\rangle}{E_{\mathbf{p}',\mathbf{k}'}-Z_{\mathbf{p},
r}-i\varepsilon},
\end{split}
\end{equation}
where $E_{\mathbf{p}',\mathbf{k}'} \equiv
E_{\mathbf{p}'}+\omega_{\mathbf{k}'}$. Substituting the
expression~\eqref{ht20} into~\eqref{ht50} for $ Z_{\mathbf{p}, r}$
we obtain
\begin{equation}\label{ht21}
\begin{split}
&Z_{\mathbf{p}, r}=E_{\mathbf{p}}~+~e^{2}\frac{m^{2}}{(2\pi)^{3}}
\int\sum\limits_{\lambda',r'}\frac{d\mathbf{p}'d\mathbf{k}'}{E_{\mathbf{p}}E_{\mathbf{p}'}2\omega_{\mathbf{k}'}}~
\delta(\mathbf{p}'-\mathbf{p}+\mathbf{k}')~\times~\\
&\times~\frac{e^{\lambda'}_{\mu}(k')\overline{u}^{r}(p)\gamma_{\mu}u^{r'}(p')
e^{\lambda'}_{\nu}(k')\overline{u}^{r'}(p')\gamma_{\nu}u^{r}(p)}{E_{\mathbf{p}',\mathbf{k}'}-Z_{\mathbf{p},
r}-i\varepsilon}~.
\end{split}
\end{equation}
Using the formal expression $\frac{1}{w\pm i\varepsilon
}\rightarrow\wp\frac{1}{w}\mp i\pi\delta(w)$ rewrite~\eqref{ht21}
in the form
\begin{align}\label{ht22}
Z_{\mathbf{p},
r}=\widetilde{E}_{\mathbf{p},r}~-~i\gamma_{\mathbf{p},r},
\end{align}
where
\begin{equation}\label{ht22}
\begin{split}
&\widetilde{E}_{\mathbf{p},r}=E_{\mathbf{p}}~+~e^{2}\frac{m^{2}}{(2\pi)^{3}}\wp
\int\sum\limits_{\lambda',r'}\frac{d\mathbf{p}'d\mathbf{k}'}{E_{\mathbf{p}}E_{\mathbf{p}'}2\omega_{\mathbf{k}'}}~
\delta(\mathbf{p}'-\mathbf{p}+\mathbf{k}')~\times~ \\
&\times~\frac{e^{\lambda'}_{\mu}(k')\overline{u}^{r}(p)\gamma_{\mu}u^{r'}(p')
e^{\lambda'}_{\nu}(k')\overline{u}^{r'}(p')\gamma_{\nu}u^{r}(p)}{E_{\mathbf{p}',\mathbf{k}'}-Z_{\mathbf{p}
r}}
\end{split}
\end{equation}
is the renormalized energy ($\wp$ stands for the principal part)
and
\begin{align}\label{ht130}
\begin{split} &\gamma_{\mathbf{p},r}~=~-
e^{2}\frac{m^{2}}{8\pi^{2}}
\int\sum\limits_{\lambda',r'}\frac{d\mathbf{p}'d\mathbf{k}'}{E_{\mathbf{p}}E_{\mathbf{p}'}2\omega_{\mathbf{k}'}}~
\delta(\mathbf{p}'-\mathbf{p}+\mathbf{k}')~\times~\\
&\times~
e^{\lambda'}_{\mu}(k')\overline{u}^{r}(p)\gamma_{\mu}u^{r'}(p')
e^{\lambda'}_{\nu}(k')\overline{u}^{r'}(p')\gamma_{\nu}u^{r}(p)
\delta(E_{\mathbf{p}',\mathbf{k}'}-Z_{\mathbf{p}, r }).
\end{split}
\end{align}
The expression~\eqref{ht21} leads to the relation
\begin{equation}\label{ht30}
\begin{split}
&E_{\mathbf{p}}-Z_{\mathbf{p}, r}~-~e^{2}\frac{m^{2}}{(2\pi)^{3}}
\int\sum\limits_{\lambda',r'}\frac{d\mathbf{p}'d\mathbf{k}'}{E_{\mathbf{p}}E_{\mathbf{p}'}2\omega_{\mathbf{k}'}}~
\delta(\mathbf{p}'-\mathbf{p}+\mathbf{k}')~\times~\\
&\times~\frac{|e^{\lambda'}_{\mu}(k')\overline{u}^{r}(p)\gamma_{\mu}u^{r'}(p')|^{2}}
{E_{\mathbf{p}',\mathbf{k}'}-Z_{\mathbf{p},r}-i\varepsilon}=0.
\end{split}
\end{equation}
From eq.~\eqref{ht30} we can obtain the connection
\begin{align}\label{ht3000}
\frac{1}{E_{\mathbf{p}',\mathbf{k}'}-Z_{\mathbf{p},r}-i\varepsilon}
=\frac{1}{E_{\mathbf{p}',\mathbf{k}'}-E_{\mathbf{p}}-i\varepsilon}
+ O(e),
\end{align}
where $O(e)$ determines the terms of higher orders on $e$.
Limiting the expression~\eqref{ht130} by order $e^{2}$ we have
\begin{equation}\label{ht301}
\begin{split}
&\gamma_{\mathbf{p},r} \approx~e^{2}\frac{m^{2}}{8\pi^{2}}
\int\sum\limits_{\lambda',r'}\frac{d\mathbf{p}'d\mathbf{k}'}{E_{\mathbf{p}}E_{\mathbf{p}'}2\omega_{\mathbf{k}'}}~
\delta(\mathbf{p}'-\mathbf{p}+\mathbf{k}')\delta(E_{\mathbf{p}',\mathbf{k}'}-E_{\mathbf{p}})
~\times~\\
&\times~|e^{\lambda'}_{\mu}(k')\overline{u}^{r}(p)\gamma_{\mu}u^{r'}(p')|^{2}~>~0,
\end{split}
\end{equation}
where for the photon $\lambda'=1,2.$ Since the
expressions~\eqref{ht30}~\eqref{ht3000} contain two complex terms
$i\gamma_{\mathbf{p}, r}$ and $i\varepsilon$, which determine a
pole in the lower half plane and a pole in the upper half plane,
the operation of integration of the expressions, which contain the
values of the form $\frac{1}
{E_{\mathbf{p}',\mathbf{k}'}-Z_{\mathbf{p},r}-i\varepsilon}$ is
accepted to determine as
follows~\cite{ppt},~\cite{pp3},~\cite{hh}: we first have to
evaluate the integration on the upper half-plane $(C^{+})$ and
then the limit of $z\rightarrow-i\gamma_{\mathbf{p}, r}$ must be
taken. For example the integration over $E_{\mathbf{p}'}$ with a
test function $g(E_{\mathbf{p}'})$ can be presented as follows
\begin{equation}\label{ht300}
\begin{split}
&\lim\limits_{z\rightarrow-i\gamma_{\mathbf{p},r}}
\Bigr{(}\int\limits_{0}^{\infty}dE_{\mathbf{p}'}\frac{g(E_{\mathbf{p}'})}{E_{\mathbf{p}'}+\omega_{\mathbf{k}'}-
\widetilde{E}_{\mathbf{p}}~-~z}\Bigl{)}_{z\in C^{+}}\equiv\\
&\equiv\int\limits_{0}^{\infty}dE_{\mathbf{p}'}\frac{g(E_{\mathbf{p}'})}{(E_{\mathbf{p}'}+\omega_{\mathbf{k}'}-
\widetilde{E}_{\mathbf{p}}~-~z)^{+}_{-i\gamma_{\mathbf{p},r}}}.
\end{split}
\end{equation}
This special feature will be used below for the determination of
the expression for the density matrix.

\section{Expression for the density matrix }

Determination of the expression~\eqref{Dts4} we will carry out for
the diagonal ($\nu$=0) matrix element of the form
$\langle\langle\gamma\gamma |\rho^{0}(t)\rangle\rangle \equiv
\rho^{0}_{\gamma\gamma}(t) $. Let initial moment of time be zero.
In this case the expression~\eqref{Dts4} can be represented as
follows
\begin{equation}\label{Dts59}
\begin{split}
 &\rho^{0}_{\gamma\gamma}(t)=\rho^{0}_{\gamma\gamma}(0)~+~
 (-i)\int\limits^{t}_{0}dt_{1}\langle\langle\gamma\gamma|\vartheta^{0}(t_{1})|\alpha\alpha\rangle
 \rangle\rho^{0}_{\alpha\alpha}(0)~+~\\
 &+~(-i)^{2}\int\limits^{t}_{0}\int\limits^{t_{1}}_{0}dt_{1}dt_{2}
 \langle\langle\gamma\gamma|\vartheta^{0}(t_{1})|\gamma '\gamma '\rangle\rangle\langle\langle \gamma '\gamma '
 |\vartheta^{0}(t_{2})|\alpha\alpha\rangle\rangle\rho^{0}_{\alpha\alpha}(0)~+~\\
&+~(-i)^{3}\int\limits^{t}_{0}\int\limits^{t_{1}}_{0}\int\limits^{t_{2}}_{0}dt_{1}dt_{2}dt_{3}
\langle\langle\gamma\gamma|\vartheta^{0}(t_{1})|\gamma'\gamma'\rangle\rangle\langle\langle
\gamma '\gamma '|\vartheta^{0}(t_{2})|\gamma
''\gamma''\rangle\rangle\times~\\
&\times\langle\langle\gamma
''\gamma''|\vartheta^{0}(t_{3})|\alpha\alpha\rangle\rangle\rho^{0}_{\alpha\alpha}(0)~+~...,~
\end{split}
\end{equation}
where summation (integration) over all internal indices $\gamma'$,
$\gamma'' $,..,$~\alpha$ is implied (It is necessary to note that
for the simplification of the expressions the normalizing volume
is implied, but it is not written.).  Eq.~\eqref{fieq} leads to
the following approximation for the collision operator
$\theta^{\nu}_{C}$
\begin{align}\label{Dts50}
\theta^{\nu}_{C}\approx L_{0}P^{\nu}~+~P^{\nu}L_{I}C^{\nu}P^{\nu},
\end{align}
where
\begin{align}\label{Dts6}
C^{\nu}=\sum\limits_{\mu\neq
\nu}P^{\mu}\frac{-1}{w^{\mu}-w^{\nu}-i\varepsilon_{\mu\nu}}L_{I}P^{\nu}.
\end{align}
From the expressions~\eqref{tet},~\eqref{Dts6} it follows the
expression for the operator $\vartheta^{\nu}(t)$
\begin{align}\label{Dts7}
\vartheta^{\nu}(t)=\sum\limits_{\mu\neq
\nu}P^{\nu}L_{I}(t)P^{\mu}\frac{-1}{w^{\mu}-w^{\nu}-i\varepsilon_{\mu\nu}}L_{I}(t)P^{\nu}.
\end{align}
Then for the $\vartheta^{0}(t)$ taking into account the condition
$d_{\mu}>d_{0}$ we obtain  ¶
\begin{align}\label{Dts8}
\vartheta^{0}(t)=\sum\limits_{\mu\neq 0}
P^{0}L_{I}(t)P^{\mu}\frac{-1}{w^{\mu}-i\varepsilon}L_{I}(t)P^{0}.
\end{align}
Using result~\eqref{Dts8} we examine the second term of
expression~\eqref{Dts59}
\begin{equation}\label{Dts9}
\begin{split}
&(-i)\int\limits^{t}_{0}dt_{1}\langle\langle\gamma\gamma|\vartheta^{0}(t_{1})|\alpha\alpha\rangle
 \rangle\rho^{0}_{\alpha\alpha}(0)~=~\\
=~&(-i)\int\limits^{t}_{0}dt_{1}\int\limits_{\alpha}\Bigl{[}\frac{\langle\gamma|H_{I}|\alpha\rangle\langle\alpha|H_{I}|\gamma\rangle}
{w^{\alpha\gamma}-i\varepsilon}~+~\frac{\langle\alpha|H_{I}|\gamma\rangle\langle\gamma|H_{I}|\alpha\rangle}
{w^{\gamma\alpha}-i\varepsilon}\Bigr{]}\rho^{0}_{\alpha\alpha}(0)~-~ \\
-~&(-i)\int\limits^{t}_{0}dt_{1}\int\limits_{\rho}\Bigl{[}\frac{\langle\gamma|H_{I}|\rho\rangle\langle\rho|H_{I}|\gamma\rangle}
{w^{\rho\gamma}-i\varepsilon}~+~\frac{\langle\rho|H_{I}|\gamma\rangle\langle\gamma|H_{I}|\rho\rangle}
{w^{\gamma\rho}-i\varepsilon}\Bigr{]}\rho^{0}_{\gamma\gamma}(0).
\end{split}
\end{equation}
Symbol $\int\limits_{\alpha,\rho}$ in~\eqref{Dts9} indicates summation over discrete and integration over continuous
variables. Selecting the state $|\gamma\rangle$ in the form $ |\mathbf{p},r\rangle$ and taking into account the
expressions~\eqref{ae90},~\eqref{ht3000},~\eqref{ht300}  we obtain
\begin{equation}\label{Dts90}
\begin{split}
&(-i)\int\limits_{\alpha}\Bigl{[}\frac{\langle\gamma|H_{I}|\alpha\rangle\langle\alpha|H_{I}|\gamma\rangle}
{w^{\alpha\gamma}-i\varepsilon}~+~\frac{\langle\alpha|H_{I}|\gamma\rangle\langle\gamma|H_{I}|\alpha\rangle}
{w^{\gamma\alpha}-i\varepsilon}\Bigr{]}\rho^{0}_{\alpha\alpha}(0)=
~2\gamma_{\mathbf{p}, r}e^{2}~\times~ \\
&\times\int\sum\limits_{\lambda',r'}d\mathbf{p}'d\mathbf{k}'\frac
{\langle
\mathbf{p},r|U|\mathbf{p}',r';~\mathbf{k}',\lambda'\rangle \langle
\mathbf{p}', r';~\mathbf{k}',\lambda'|U|\mathbf{p},r\rangle}
{(E_{\mathbf{p}',\mathbf{k}'}-
\widetilde{E}_{\mathbf{p}}~-~z)^{+}_{-i\gamma_{\mathbf{p},
r}}(E_{\mathbf{p}',\mathbf{k}'}-
\widetilde{E}_{\mathbf{p}}~-~z)^{-}_{+i\gamma_{\mathbf{p}, r}}}
~\times~\\
&\times\rho^{0}_{\mathbf{p}',r';\mathbf{k}',\lambda'~\mathbf{p}',r';\mathbf{k}',\lambda'}(0),
\end{split}
\end{equation}
where $U=\int
N(\overline{\psi}(x)\gamma_{\mu}\psi(x))A_{\mu}(x)d\mathbf{x}$ and
$\gamma_{\mathbf{p}, r}$ is determined~\eqref{ht130}. The
designation $(E_{\mathbf{p}',\mathbf{k}'}-
\widetilde{E}_{\mathbf{p}}~-~z)^{-}_{+i\gamma_{\mathbf{p}, r}}$
corresponds to the integration, which first of all is carried out
in the lower half complex plane $(C^{-})$ and, after that, the
limit of $z\rightarrow +i\gamma_{\mathbf{p}, r}$ is taken. \\
For convenience of the further consideration let me introduce the
new designations. We define the function
\begin{align}\label{Dts900}
\Gamma_{\tau
i}\equiv2\gamma_{\tau}e^{2}\frac{\langle\tau|U|i\rangle\langle
i|U|\tau\rangle}{(E_{i}-\widetilde{E}_{\tau}-z)^{+}_{-i\gamma_{\tau}}
(E_{i}-\widetilde{E}_{\tau}-z)^{-}_{+i\gamma_{\tau}}},
\end{align}
where Greek and Roman indices  $|\tau\rangle$, $|i\rangle$
correspond to the one-particle and two-particles states,
respectively. For the function $\Gamma_{\tau i}$ it is  possible
to determine the rules
\begin{equation}\label{Dts1900}
\begin{split}
&\Gamma_{\tau
i}\rho^{0}_{ii}(0)\equiv2\gamma_{\tau}e^{2}\int\limits_{i}\frac{\langle\tau|U|i\rangle\langle
i|U|\tau\rangle}{(E_{i}-\widetilde{E}_{\tau}-z)^{+}_{-i\gamma_{\tau}}
(E_{i}-\widetilde{E}_{\tau}-z)^{-}_{+i\gamma_{\tau}}}\rho^{0}_{ii}(0),\\
&\Gamma_{\tau i}\Gamma_{\beta i}\rho^{0}_{ii}(0)
\equiv2\gamma_{\tau}e^{2}\int\limits_{i}\frac{\langle\tau|U|i\rangle\langle
i|U|\tau\rangle}{(E_{i}-\widetilde{E}_{\tau}-z)^{+}_{-i\gamma_{\tau}}
(E_{i}-\widetilde{E}_{\tau}-z)^{-}_{+i\gamma_{\tau}}}~\times~\\
&\times~2\gamma_{\beta}e^{2}\frac{\langle\beta|U|i\rangle\langle
i|U|\beta\rangle}{(E_{i}-\widetilde{E}_{\beta}-z)^{+}_{-i\gamma_{\beta}}
(E_{i}-\widetilde{E}_{\beta}-z)^{-}_{+i\gamma_{\beta}}}\rho^{0}_{ii}(0).
\end{split}
\end{equation}
The rules~\eqref{Dts1900} lead to the following expression
\begin{equation}\label{Dts190}
\begin{split}
&(-i)\int\limits_{\alpha}\Bigl{[}\frac{\langle\gamma|H_{I}|\alpha\rangle\langle\alpha|H_{I}|\gamma\rangle}
{w^{\alpha\gamma}-i\varepsilon}~+~\frac{\langle\alpha|H_{I}|\gamma\rangle\langle\gamma|H_{I}|\alpha\rangle}
{w^{\gamma\alpha}-i\varepsilon}\Bigr{]}\rho^{0}_{\alpha\alpha}(0)=\\
&=\Gamma_{\mathbf{p}, r~\mathbf{p}',r';\mathbf{k}',\lambda'}
\rho^{0}_{\mathbf{p}',r';\mathbf{k}',\lambda'~\mathbf{p}',r';\mathbf{k}',\lambda'}(0).
\end{split}
\end{equation}
Substituting~\eqref{psidecoper50},~\eqref{ae},~\eqref{ae90},~\eqref{ht3000},~\eqref{ht300}
in the second term of expression~\eqref{Dts9} we obtain
\begin{equation}\label{gh}
\begin{split}
&(-i)\int\limits_{\rho}\Bigl{[}\frac{\langle\gamma|H_{I}|\rho\rangle\langle\rho|H_{I}|\gamma\rangle}
{w^{\rho\gamma}-i\varepsilon}~+~\frac{\langle\rho|H_{I}|\gamma\rangle\langle\gamma|H_{I}|\rho\rangle}
{w^{\gamma\rho}-i\varepsilon}\Bigr{]}\rho^{0}_{\gamma\gamma}(0)=
\\
&=2\gamma_{{\mathbf{p}, r}}\rho^{0}_{\mathbf{p},
r~\mathbf{p},r}(0).
\end{split}
\end{equation}
In expression~\eqref{gh} integration (summation) over the state
$|\mathbf{p}, r\rangle$ is not carried out. Finally we obtain the
result
\begin{equation}\label{fo}
\begin{split}
&(-i)\int\limits^{t}_{0}dt_{1}\langle\langle\gamma\gamma|\vartheta^{0}(t_{1})|\alpha\alpha\rangle
 \rangle\rho^{0}_{\alpha\alpha}(0)=(\Gamma_{\mathbf{p},
r~\mathbf{p}',r';\mathbf{k}',\lambda'}~\times~\\
&\times~\rho^{0}_{\mathbf{p}',r';\mathbf{k}',\lambda'~\mathbf{p}',r';\mathbf{k}',\lambda'}(0)-2\gamma_{{\mathbf{p},
r}}\rho^{0}_{\mathbf{p}, r~\mathbf{p}, r}(0))t.
\end{split}
\end{equation}
Analogously for the the third and fourth contributions to the
expression~\eqref{Dts59} we obtain
\begin{equation}\label{fo}
\begin{split}
&(-i)^{2}\int\limits^{t}_{0}\int\limits^{t_{1}}_{0}dt_{1}dt_{2}
 \langle\langle\gamma\gamma|\vartheta^{0}(t_{1})|\gamma '\gamma '\rangle\rangle\langle\langle \gamma '\gamma '
 |\vartheta^{0}(t_{2})|\alpha\alpha\rangle\rangle\rho^{0}_{\alpha\alpha}(0)~=~
 \\
 &=~((2\gamma_{{\mathbf{p}, r}})^{2}\rho^{0}_{\mathbf{p}, r~\mathbf{p}, r}(0)-
2\gamma_{{\mathbf{p}, r}}\Gamma_{\mathbf{p},
r~\mathbf{p}',r';\mathbf{k}',\lambda'}\rho^{0}_{\mathbf{p}',r';\mathbf{k}',\lambda'~\mathbf{p}',r';\mathbf{k}',\lambda'
}(0)~+\\
& + \Gamma_{\mathbf{p},
r~\mathbf{p}',r';\mathbf{k}',\lambda'}\Gamma_{\mathbf{p}'', r''
~\mathbf{p}',r';\mathbf{k}',\lambda'}\rho^{0}_{\mathbf{p}'', r'';
~\mathbf{p}'', r''}(0)~-~
\\&- \Gamma_{\mathbf{p}, r~\mathbf{p}',r';\mathbf{k}',\lambda'}
\int\sum\limits_{r''}d\mathbf{p}''\Gamma_{\mathbf{p}'',
r''~\mathbf{p}',r';\mathbf{k}',\lambda'}\rho^{0}_{\mathbf{p}',r';\mathbf{k}',\lambda'~\mathbf{p}',r';\mathbf{k}',\lambda'}(0))\frac{t^{2}}{2!},
\end{split}
\end{equation}
\begin{equation}\label{fo2}
\begin{split}
&(-i)^{3}\int\limits^{t}_{0}\int\limits^{t_{1}}_{0}\int\limits^{t_{2}}_{0}dt_{1}dt_{2}dt_{3}
\langle\langle\gamma\gamma|\vartheta^{0}(t_{1})|\gamma'\gamma'\rangle\rangle\langle\langle
\gamma '\gamma '|\vartheta^{0}(t_{2})|\gamma
''\gamma''\rangle\rangle~\times~\\
&\times\langle\langle\gamma
''\gamma''|\vartheta^{0}(t_{3})|\alpha\alpha\rangle\rangle
\rho^{0}_{\alpha\alpha}(0)= (-(2\gamma_{{\mathbf{p},
r}})^{3}\rho^{0}_{\mathbf{p}, r~\mathbf{p}, r}(0)~+~\\
&+~(2\gamma_{{\mathbf{p}, r}})^{2}\Gamma_{\mathbf{p},
r~\mathbf{p}',r';\mathbf{k}',\lambda'}
\rho^{0}_{\mathbf{p}',r';\mathbf{k}',\lambda'~\mathbf{p}',r';\mathbf{k}',\lambda'}(0)+
2\gamma_{\mathbf{p}, r}\Gamma_{\mathbf{p}, r~\mathbf{p}',r';\mathbf{k}',\lambda'}~\times \\
&\times\int\sum\limits_{r''}d\mathbf{p}''\Gamma_{\mathbf{p}'',
r''~\mathbf{p}',r';\mathbf{k}',\lambda'}
\rho^{0}_{\mathbf{p}',r';\mathbf{k}',\lambda'~\mathbf{p}',r';\mathbf{k}',\lambda'}(0)+
\Gamma_{\mathbf{p}, r~\mathbf{p}',r';\mathbf{k}',\lambda'}~\times\\
&\times\Gamma_{\mathbf{p}'',
r''~\mathbf{p}',r';\mathbf{k}',\lambda'} \Gamma_{\mathbf{p}'',
r''~\mathbf{p}''',r''';\mathbf{k}''',\lambda'''}\rho^{0}_{\mathbf{p}''',r''';\mathbf{k}''',\lambda'''~
\mathbf{p}''',r''';\mathbf{k}''',\lambda'''}~-\\
&-\Gamma_{\mathbf{p},
r~\mathbf{p}',r';\mathbf{k}',\lambda'}\Gamma_{\mathbf{p}'',
r''~\mathbf{p}',r';\mathbf{k}',\lambda'}2\gamma_{\mathbf{p}'', r''}~\times\\
&\times\rho^{0}_{\mathbf{p}'', r''~\mathbf{p}'', r''}(0)-
\Gamma_{\mathbf{p},
r~\mathbf{p}',r';\mathbf{k}',\lambda'}\int\sum\limits_{r''}d\mathbf{p}''
\Gamma_{\mathbf{p}'', r''~\mathbf{p}',r';\mathbf{k}',\lambda'}~\times \\
&\times\Gamma_{\mathbf{p}''',
r'''~\mathbf{p}',r';\mathbf{k}',\lambda'} \rho^{0}_{\mathbf{p}''',
r'''~\mathbf{p}''', r'''}(0)+ \Gamma_{\mathbf{p},
r~\mathbf{p}',r';\mathbf{k}',\lambda'}\int\sum\limits_{r''}d\mathbf{p}''
~\times \\
&\times\Gamma_{\mathbf{p}'',
r''~\mathbf{p}',r';\mathbf{k}',\lambda'}
\int\sum\limits_{r'''}d\mathbf{p}'''\Gamma_{\mathbf{p}''',
r'''~\mathbf{p}',r';\mathbf{k}',\lambda'}\rho^{0}_{\mathbf{p}',r';\mathbf{k}',\lambda'
~\mathbf{p}',r';\mathbf{k}',\lambda'}(0)~-~\\
&-~2\gamma_{\mathbf{p}, r}\Gamma_{\mathbf{p},
r~\mathbf{p}',r';\mathbf{k}',\lambda'}\Gamma_{\mathbf{p}'',
r''~\mathbf{p}',r';\mathbf{k}',\lambda'}\rho^{0}_{\mathbf{p}'',
r''~\mathbf{p}'', r''}(0))\frac{t^{3}}{3!},
\end{split}
\end{equation}
where the procedures of integration and summing are achieved on
all continuous and discrete repeating indices besides the indices
$\mathbf{p}, r$ which correspond to the state $|\gamma\rangle$.\\
Studies of the expression~\eqref{Dts59} lead to the genealogical
connections, where each of the foregoing contribution gives birth
to the following contribution which determines the sequential term
of the sum. For example, contribution
$-2\gamma_{\gamma}\rho^{0}_{\gamma\gamma}(0)$, which determines
second term in the expression~\eqref{Dts59}, is the ancestor of
the contributions
$(2\gamma_{\gamma})^{2}\rho^{0}_{\gamma\gamma}(0)$,
$-2\gamma_{\gamma}\Gamma_{\gamma i}\rho^{0}_{ii}(0)$,
$-(2\gamma_{\gamma})^{3}\rho^{0}_{\gamma\gamma}(0)$,
$(2\gamma_{\gamma})^{2}\Gamma_{\gamma i}\rho^{0}_{ii}(0),$
...,$(2\gamma_{\gamma})^{n}\rho^{0}_{\gamma\gamma}(0)$,
$-(2\gamma_{\gamma})^{n-1}\Gamma_{\gamma i}\rho^{0}_{ii}(0),$ ...
. Analogously, contribution $\Gamma_{\gamma i}\rho^{0}_{ii}(0)$,
determining second term generates contributions: $\Gamma_{\gamma
i}\Gamma_{\beta i}\rho^{0}_{\beta\beta}(0)$, $-\Gamma_{\gamma
i}\int\limits_{\rho}\Gamma_{\rho i}\rho^{0}_{ii}(0)$ and so on.
For example, for the fifth order we have the connections  \\
\begin{bundle}{$-(2\gamma_{\gamma})^{5}\rho^{0}_{\gamma\gamma}$}
\chunk{$(2\gamma_{\gamma})^{6}\rho^{0}_{\gamma\gamma}$}
~~\chunk{$-(2\gamma_{\gamma})^{5}\Gamma_{\gamma i}\rho^{0}_{ii}$}
\end{bundle}
~~\begin{bundle}{$-(2\gamma_{\gamma})^{5}\Gamma_{\gamma
i}\rho^{0}_{ii}$} \chunk{$-(2\gamma_{\gamma})^{5}\Gamma_{\gamma
i}\Gamma_{\beta i}\rho^{0}_{\beta\beta}$}
~~~\chunk{$(2\gamma_{\gamma})^{5}\Gamma_{\gamma i
}\int\limits_{\rho}\Gamma_{\rho i }\rho^{0}_{ii}$}
\end{bundle}\\
\begin{bundle}{$-(2\gamma_{\gamma})^{5}\Gamma_{\gamma
i}\Gamma_{\beta i}\rho^{0}_{\beta\beta}$}
\chunk{$(2\gamma_{\gamma})^{5}\Gamma_{\gamma
i}2\gamma_{\beta}\Gamma_{\beta i}\rho^{0}_{\beta\beta}$}
~~~\chunk{$-(2\gamma_{\gamma})^{5}\Gamma_{\gamma i }\Gamma_{\beta
i }\Gamma_{\beta k }\rho^{0}_{kk}$}
\end{bundle}
\\

where integration and summing are achieved on all repeating
indices besides $\gamma$. Further analysis of the terms of the
expression~\eqref{Dts59}, leads to the extremely great variety of
contributions of higher order on $e$. I limit my analysis by the
contributions determining the structure of the density matrix in
the approximate form
\begin{equation}\label{fo4}
\begin{split}
&\rho^{0}_{\mathbf{p}, r~\mathbf{p}, r}(t)\approx
e^{-2\gamma_{\mathbf{p}, r} t}\rho^{0}_{\mathbf{p}, r~\mathbf{p},
r}(0) + (1-e^{-2\gamma_{\mathbf{p}, r} t})\Gamma_{\mathbf{p},
r~\mathbf{p}',r';\mathbf{k}',\lambda' }~\times\\
&\times\rho^{0}_{\mathbf{p}',r';\mathbf{k}',\lambda'~
\mathbf{p}',r';\mathbf{k}',\lambda' }(0).
\end{split}
\end{equation}
The expression~\eqref{fo4} is determined so that the function
$\Gamma_{\mathbf{p}, r~\mathbf{p}',r';\mathbf{k}',\lambda' }$ does
not contain the value $2\gamma_{\mathbf{p}, r}$.
Expression~\eqref{fo4} follows from equation~\eqref{kineq} and
corresponds to the kinetic, irreversible evolution of the unstable
electromagnetic system in the time to the equilibrium state. Thus,
the ordering in the time leads to the complex eigenvalues. Such
complex eigenvalues make it possible to describe the relaxation
process in other words the irreversible process without appearance
of the other spontaneous, unstable states.

\section{Numerical calculation}

I examine the first term of the expression~\eqref{fo4} which is
the probability of finding Fermi - Dirac particle with momentum
$\mathbf{p}$ and helicity $r$ depending on the time. We will
assume that the particle is not polarized. The averaging over $r$
leads to the following result
\begin{align}\label{gamma}
\begin{split}
\frac{1}{2}\sum\limits_{r}e^{-2\gamma_{\mathbf{p}, r}
t}\rho^{0}_{\mathbf{p}, r~\mathbf{p},
r}(0)=e^{-2\gamma_{\mathbf{p}, r=\pm 1}
t}\frac{1}{2}\sum\limits_{r}\rho^{0}_{\mathbf{p}, r~\mathbf{p},
r}(0)\equiv {\rho}^{0}_{\mathbf{p}~\mathbf{p}}(t),
\end{split}
\end{align}
where
\begin{align}\label{gamma}
\begin{split}
&\gamma_{\mathbf{p},r=\pm 1} \approx~- e^{2}\frac{m^{2}}{8\pi^{2}}
\int\sum\limits_{\lambda'}\frac{d\mathbf{p}'d\mathbf{k}'}{E_{\mathbf{p}}E_{\mathbf{p}'}2\omega_{\mathbf{k}'}}~
\delta(\mathbf{p}'-\mathbf{p}+\mathbf{k}')~\times\\
&\times~\delta(E_{\mathbf{p}',\mathbf{k}'}-E_{\mathbf{p}})
\frac{1}{2}Tr(\mathbf{e}^{\lambda'}(k')\cdot\bgamma\Lambda(p')
\mathbf{e}^{\lambda'}(k')\cdot\bgamma\Lambda(p)) \\
&\text{and}~ \Lambda(p)=\frac{ \hat{p}+im}{2im}.
\end{split}
\end{align}
Using the expression~\eqref{pca3} we represent the density matrix
$\rho^{0}_{\mathbf{p}, r~\mathbf{p}, r}(0)$ in the form
\begin{align}\label{fo40}
\rho^{0}_{\mathbf{p}, r~\mathbf{p}, r}(0)=\rho_{\mathbf{p},
r~\mathbf{p}, r}(0)-(D^{0}C^{0}\rho(0))_{\mathbf{p}, r~\mathbf{p},
r}.
\end{align}
From the expression~\eqref{Dts6} and relation
$D^{\nu}=(C^{\nu})^{\ast}$~\cite{opp} we can find
\begin{align}\label{fo41}
\begin{split}
&\rho^{0}_{\mathbf{p}, r~\mathbf{p}, r}(0)\approx\rho_{\mathbf{p},
r~\mathbf{p}, r}(0)-e^{2}\frac{m^{2}}{(2\pi)^{3}}
\int\sum\limits_{\lambda',r'}\frac{d\mathbf{p}'d\mathbf{k}'}{E_{\mathbf{p}}E_{\mathbf{p}'}2\omega_{\mathbf{k}'}}~
\delta(\mathbf{p}'-\mathbf{p}+\mathbf{k}')~\times~\\
&\times~\Bigl{(}\frac{|e^{\lambda'}_{\mu}(k')\overline{u}^{r}(p)\gamma_{\mu}u^{r'}(p')|^{2}}
{(E_{\mathbf{p}',\mathbf{k}'}-Z_{\mathbf{p},
r}-i\varepsilon)^{2}}+ c.c.\Bigl{)}\rho_{\mathbf{p}, r~\mathbf{p},
r}(0),
\end{split}
\end{align}
where $c.c.$ means the complex conjugate. Density matrix
$\rho_{\mathbf{p}, r~\mathbf{p}, r}(0)$ has the form~\cite{Bil}
\begin{align}\label{rel}
\rho_{\mathbf{p}, r~\mathbf{p}, r}(0)=u^{r}(p)\overline{u}^{r}(p).
\end{align}
We determine the expression for
${\rho}^{0}_{\mathbf{p}~\mathbf{p}}(t)$ being limited to term of
lowerst order on $e$. In this case we have
\begin{align}\label{rel2}
{\rho}^{0}_{\mathbf{p}~\mathbf{p}}(t)\approx
e^{-2\gamma_{\mathbf{p}, r=\pm 1}
t}{\rho}_{\mathbf{p}~\mathbf{p}}(0)
\end{align}
with
\begin{align}\label{rel23}
{\rho}_{\mathbf{p}~\mathbf{p}}(0)=\frac{1}{2}\sum\limits_{r}u^{r}(p)\overline{u}^{r}(p)=\frac{1}{2}\Lambda(p)
{~\text -}
\end{align}
relativistic density matrix of the Fermi-Dirac particle. Let
estimate the density matrix summing up the diagonal elements of
the expression~\eqref{rel2}. This procedure results into
\begin{align}\label{gdiag}
({\rho}^{0}_{\mathbf{p}~\mathbf{p}}(t))_{diag}\approx
e^{-2\gamma_{\mathbf{p}, r=\pm 1}
t}({\rho}_{\mathbf{p}~\mathbf{p}}(0))_{diag},~
{\text{where}~({\rho}_{\mathbf{p}~\mathbf{p}}(0))_{diag}=1.}
\end{align}
Since $\gamma_{\mathbf{p},r=\pm 1}$ depends on the momentum
$\mathbf{p}$ we examine the special case, when the angle
$\vartheta_{\mathbf{p}}$ of vector $\mathbf{p}$ (in spherical
coordinates) is zero. Summation over $\lambda'$ and integration
over $\delta$ - functions give the result
\begin{align}\label{rel2348}
\begin{split}
\gamma_{\mathbf{\mid p\mid},r=\pm
1}=\frac{\alpha}{2\pi}(I_{1}+I_{2}+I_{3}+I_{4}),
\end{split}
\end{align}
where
\begin{align}\label{rel234w}
\begin{split}
I_{1}=-\frac{\omega|\mathbf{p}|^{2}}{2E_{\mathbf{p}}}
\int\frac{\cos^{2}(\vartheta_{\mathbf{k}'})d\Omega_{\mathbf{k}'}}
{(|\mathbf{p}|^{2}+\omega^{2}+m^{2}-2|\mathbf{p}|\omega
\cos(\vartheta_{\mathbf{k}'}))^{1/2}+\omega-|\mathbf{p}|\cos(\vartheta_{\mathbf{k}'})},
\end{split}
\end{align}
\begin{align}\label{rel2345}
\begin{split}
I_{2}=\frac{\omega^{2}|\mathbf{p}|}{2E_{\mathbf{p}}}
\int\frac{\cos(\vartheta_{\mathbf{k}'})d\Omega_{\mathbf{k}'}}
{(|\mathbf{p}|^{2}+\omega^{2}+m^{2}-2|\mathbf{p}|\omega
\cos(\vartheta_{\mathbf{k}'}))^{1/2}+\omega-|\mathbf{p}|\cos(\vartheta_{\mathbf{k}'})},
\end{split}
\end{align}
\begin{align}\label{rel2346}
\begin{split}
I_{3}=\frac{\omega}{2}
\int\frac{(|\mathbf{p}|^{2}+\omega^{2}+m^{2}-2|\mathbf{p}|\omega
\cos(\vartheta_{\mathbf{k}'}))^{1/2}d\Omega_{\mathbf{k}'}}
{(|\mathbf{p}|^{2}+\omega^{2}+m^{2}-2|\mathbf{p}|\omega
\cos(\vartheta_{\mathbf{k}'}))^{1/2}+\omega-|\mathbf{p}|\cos(\vartheta_{\mathbf{k}'})},
\end{split}
\end{align}
\begin{align}\label{rel2347}
\begin{split}
I_{4}=-\frac{\omega m^{2}}{2E_{\mathbf{p}}}
\int\frac{d\Omega_{\mathbf{k}'}}
{(|\mathbf{p}|^{2}+\omega^{2}+m^{2}-2|\mathbf{p}|\omega
\cos(\vartheta_{\mathbf{k}'}))^{1/2}+\omega-|\mathbf{p}|\cos(\vartheta_{\mathbf{k}'})}
\end{split}
\end{align}
and $\omega$ is the energy of photon (without the Doppler effect. In our case the Doppler effect is not significant).
The calculation of the expression~\eqref{gdiag} was accomplished numerically with the use of program Mathematica and
the following approximation
\begin{align}\label{rel234}
\begin{split}
(|\mathbf{p}|^{2}+\omega^{2}+m^{2}-2|\mathbf{p}|\omega
\cos(\vartheta_{\mathbf{k}'}))^{1/2}\approx\frac{|\mathbf{p}|^{2}+\omega^{2}+m^{2}-
|\mathbf{p}|\omega\cos(\vartheta_{\mathbf{k}'})}{(|\mathbf{p}|^{2}+\omega^{2}+m^{2})^{1/2}}.
\end{split}
\end{align}
The calculated results are represented in figs.1-3, where the
process like $e\rightarrow e+\gamma$ (bremsstrahlung of electron)
is examined. The calculations are executed for the different
values of momentum $|\mathbf{p}|$ of electron. It is seen, the
time evolution of the density matrix depends on the value of
momentum of particle: the density matrix is decreased with an
increasing of momentum
$|\mathbf{p}|$. \\
It is necessary to note that these calculations do not determine
the physical bremsstrahlung. First of all because the used model
of interaction is determined for the bare not dressed particles.
Furthermore it is known that electron in the bare, free state
cannot either absorb or radiate the photons. Strictly speaking,
the energy $\omega$ of the radiated photon in the
expressions~\eqref{rel234w}~-~\eqref{rel2347} must be zero.
Nevertheless, in the model it is assumed that electron interacts
with the external electromagnetic field. Therefore, the energy
$\omega$ of the radiated photon was forced different from zero.
The model  makes it possible to develop the procedure for the
description of the realizable irreversible processes. In the
work~\cite{shirm} in the framework of Prigogine's principles the
weak interaction like $\pi^{\pm}$ - meson decay is investigated.

\section{Concluding remarks}

Let me briefly summarize the results. The time irreversible
evolution of the relativistic, unstable electromagnetic system is
investigated in the framework of Prigogine's principles of
description of nonequilibrium states on the basis of unified
formulation of quantum and kinetic dynamics. As a result the
expression for the density matrix determining irreversible
evolution of the relativistic unstable system in the time was
obtained. Although I do not examine the question of the
determination of observed physical process, the approach makes it
possible to define the expression, which can be initial for the
further construction of the irreversible relativistic model of
time evolution of the physical relativistic unstable systems. It
is interesting to investigate the possibility of applying the
developed procedure for the time irreversible description of the
observed physical processes such as relaxation of the unstable
states of atoms and atomic nuclei, bremsstrahlung, particle decay.
All these problems are very debatable and require further
consideration.

\vspace*{8mm}

{\bf Acknowledgements}

I am grateful to Dr.A.A.~Goy for the helpful suggestions and
Dr.A.V.
Molochkov, Dr.D.V.~Shulga for the support of this work. The
work was written with the support of the State Education Institute
of the Higher Vocational Education "Russian custom academy"
Vladivostok branch.

\newpage
\vspace*{5mm}
\begin{figure}[h]
\begin{center}
\begin{minipage}[b]{.8\textwidth}
\hspace*{10mm}
\includegraphics[width=0.8\textwidth]{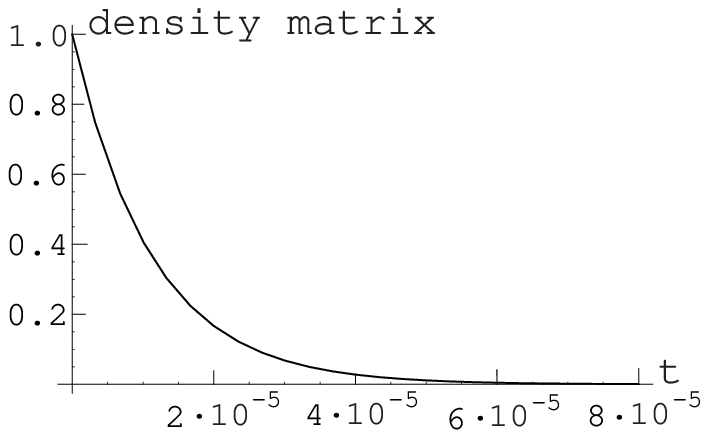}
\vspace*{1mm}\caption{\label{Rasp_ud}Density matrix
$({\rho}^{0}_{\mathbf{p}~\mathbf{p}}(t))_{diag}:~m=0.51~MeV,~
|\textbf{p}|=0~MeV,~\omega=12.8~eV$, t(sec.)}
\end{minipage}
\end{center}
\end{figure}

\vspace*{5mm}
\begin{figure}[h]
\begin{center}
\begin{minipage}[b]{.8\textwidth}
\hspace*{10mm}
\includegraphics[width=0.8\textwidth]{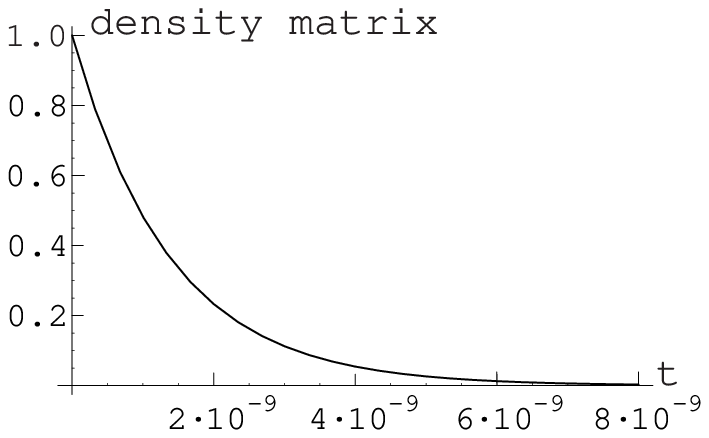}
\vspace*{1mm}\caption{\label{Rasp_ud}Density matrix
$({\rho}^{0}_{\mathbf{p}~\mathbf{p}}(t))_{diag}:~m=0.51~MeV,~
|\textbf{p}|=0.001~MeV,~\omega=12.8~eV$, t(sec.)}
\end{minipage}
\end{center}
\end{figure}

\vspace*{5mm}
\begin{figure}[h]
\begin{center}
\begin{minipage}[b]{.8\textwidth}
\hspace*{10mm}
\includegraphics[width=0.8\textwidth]{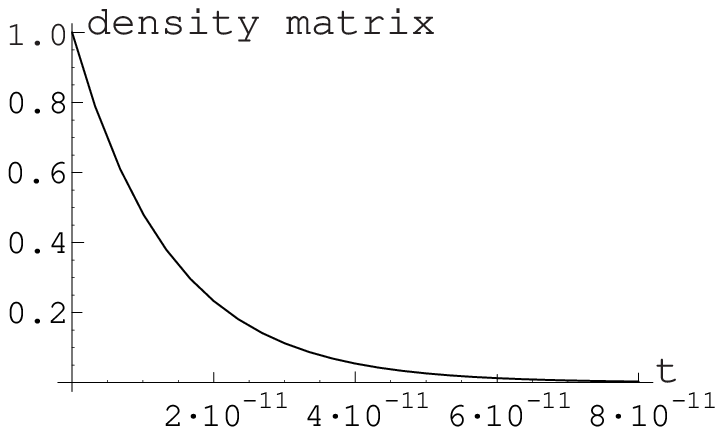}
\vspace*{1mm}\caption{\label{Rasp_ud}Density matrix
$({\rho}^{0}_{\mathbf{p}~\mathbf{p}}(t))_{diag}:~m=0.51~MeV,~
|\textbf{p}|=0.01~MeV,~\omega=12.8~eV$, t(sec.)}
\end{minipage}
\end{center}
\end{figure}
\end{document}